\documentclass[twocolumn,showpacs,floats,floatfix,superscriptaddress,aps,pra]{revtex4}

\usepackage{amsfonts}
\usepackage{amssymb}
\usepackage{amsmath}
\usepackage{graphicx}
\usepackage{bm}
\usepackage{color}
\usepackage{epsfig}
\usepackage{ifthen}

\def\bM{\left[ \matrix}
\def\eM{ \right]}
\def\bM{\begin{matrix}}
\def\eM{\end{matrix}}

\def\Black{}

\def\ignore#1{}
\def\be{\begin{equation}}
\def\ee{\end{equation}}
\def\bea{\begin{eqnarray}}
\def\eea{\end{eqnarray}}
\def\bse{\begin{subequations}}
\def\ese{\end{subequations}}
\def\bc{\begin{center}}
\def\ec{\end{center}}

\def\C{\mathbf{C}}

\def\tC{\widetilde{\mathbf{C}}}

\def\tDelta{\widetilde{\Delta}}

\def\R{ \mathsf{R} }
\def\t{}

\def\tV{\widetilde{\Omega}}

\def\W{\mathsf{W}}
\def\tW{\widetilde{\mathsf{W}}}

\def\tpsi{\widetilde{\psi}}
\def\state#1{\psi_{#1}}
\def\tstate#1{\tpsi_{#1}}

\def\state#1{\psi_{#1}}

\newlength{\frankwidth}
\begin{document} 

\author{A. A. Rangelov}
\affiliation{Fachbereich Physik der Universit\"{a}t, Erwin-Schr\"{o}dinger-Str., 67653 Kaiserslautern, Germany}
\affiliation{Permanent address: Department of Physics, Sofia University, James Bourchier 5 blvd., 1164 Sofia, Bulgaria}

\author{N. V. Vitanov}
\affiliation{Department of Physics, Sofia University, James Bourchier 5 blvd., 1164 Sofia, Bulgaria}
\affiliation{Institute of Solid State Physics, Bulgarian Academy of Sciences, Tsarigradsko chauss\'{e}e 72, 1784 Sofia, Bulgaria}

\author{B. W. Shore}
\affiliation{Fachbereich Physik der Universit\"{a}t, Erwin-Schr\"{o}dinger-Str., 67653 Kaiserslautern, Germany}
\altaffiliation{Permanent address: 618 Escondido Cir., Livermore, CA}

\title{Population trapping in three-state quantum loops revealed by Householder reflections}

\date{\today}

\begin{abstract}
Population trapping occurs when a particular quantum-state superposition is immune to action by a specific interaction, such as the well-known dark state in a three-state lambda system.
We here show that in a three-state loop linkage, a Hilbert-space Householder reflection breaks the loop and presents the linkage as a single chain.
With certain conditions on the interaction parameters, this chain can break into a simple two-state system and an additional spectator state.
Alternatively, a two-photon resonance condition in this Householder-basis chain can be enforced, which heralds the existence of another spectator state.
These spectator states generalize the usual dark state to include contributions from all three bare basis states and disclose hidden population trapping effects, and hence hidden constants of motion.
Insofar as a spectator state simplifies the overall dynamics, its existence facilitates the derivation of analytic solutions and the design of recipes for quantum state engineering in the loop system.
Moreover, it is shown that a suitable sequence of Householder transformations can cast an arbitrary $N$-dimensional hermitian Hamiltonian into a tridiagonal form.
The implication is that a general $N$-state system, with arbitrary linkage patterns where each state connects to any other state, can be reduced
 to an equivalent chainwise-connected system, with nearest-neighbor interactions only,
 with ensuing possibilities for discovering hidden multidimensional spectator states and constants of motion. \Black
\end{abstract}

 \maketitle

\section{Introduction}

Descriptions of optical excitation of few-state quantum systems traditionally make use of the rotating-wave approximation (RWA),
 in which the Hilbert-space unit vectors (the bare quantum states) rotate with angular velocities that are fixed at various laser carrier frequencies, and the Hamiltonian,
 with the neglect of rapidly varying terms, becomes a  matrix of slowly varying Rabi frequencies and detunings \cite{All75,Shore}.
For three states the usual electric-dipole selection rules of optical transitions produce a simple chain of inter-state linkages, depicted as either a ladder, a lambda or a vee.

For some time it has been known that, either by means of a rotation of the arbitrary quantization axis for defining magnetic sublevels,
 or by more general reorganization of the Hilbert-state basis states (a Morris-Shore transformation \cite{Morris83}),
 such patterns can be presented as a set of independent two-state excitations (bright states) together with spectator states that are unaffected by the specific radiation (dark states or spectator states).

\begin{figure}[tb]
\epsfig{width=36mm,file=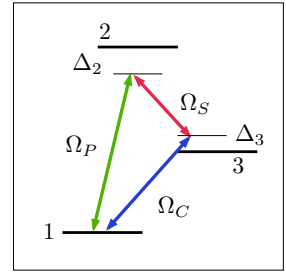} \caption{(Color online) RWA
linkage pattern for a loop, showing linkages: states 1 and 2 by
Rabi frequency $\Omega_P$, states 2 and 3 by $\Omega_S$ and states
1 and 3 by $\Omega_C$. The energy levels are shown with an
ordering appropriate to state $\state{1}$ as ground state and
state $\state{2}$ as excited state, but the symmetry of the loop
linkage allows initial population in any state.} \label{fig-loop3}
\end{figure}

The presence of a third interaction, linking the two states that terminate the three-state chain, turns the linkage pattern into a loop, see Fig. \ref{fig-loop3}.
Such interaction would violate the usual selection rules for electric-dipole radiation (which connects only states of opposite parity),
 but is possible for a variety of other interactions, such as occur with two-photon optical transitions or microwave transitions between hyperfine states.
To avoid the presence of rapidly varying exponential phases in the RWA Hamiltonian, such a link should occur with carrier frequency suitably chosen.

Within the RWA there is no longer a distinction of the original bare-state energies; all that matters is the detunings, i.e. differences between a Bohr frequency and the associated laser-field carrier frequency.
Nonetheless, it is traditional, when depicting the linkage pattern of laser-induced interactions, to place representations of the states in a vertical direction
 ranked according to the original bare-state energies, and to consider excitation in which the initial population resides entirely in a ground state.
Such display convention is particularly useful in emphasizing the difference between low-energy stable or metastable states, unable to radiatively decay, and excited states, from which spontaneous emission is possible -- visible as fluorescence.

The three-state loop is the simplest example of discrete quantum states that can exhibit nontrivial probability-amplitude interferences, and hence it has attracted continuing interest
\cite{Buckle86,Krinitzky86,Carroll88,Kosachiov92,Maichen95,Maichen96,Unanyan97,Fleischhauer99,Shapiro,Liu,Metz06}.

With the loop pattern it is not immediately obvious that any simple restructuring of the Hilbert-space coordinates will produce a single unit vector that is immune to the effects of any interaction -- a spectator state.
For example, the loop system does not satisfy an essential  condition  for the MS transformation \cite{Morris83}, namely that the quantum states be classified in two sets, with transitions only between the sets, not within them.

A special case of the three-state loop was considered by Carroll and Hioe \cite{Carroll88}.
They presented analytical solutions for the probability amplitudes when three resonant laser pulses of different shapes were present and two of the couplings where real, while the third was purely imaginary.
For this special case, the underlying SU(2) symmetry allows the three-state loop to be reduced to an effective two-state system.

Another fully resonant three-state loop was examined by Unanyan \textit{et al.} \cite{Unanyan97}.
In that work  a  pulsed quasistatic magnetic field supplemented the two optical pulses of a lambda linkage used for stimulated Raman adiabatic passage (STIRAP) \cite{STIRAP}.
This additional field provided a supplement to the usual adiabatic constraints and allowed a reduction of diabatic loss, thereby improving the usual adiabatic constraints on achieving complete population transfer.

The three-state loop was examined also by Fleischhauer \textit{et al.} \cite{Fleischhauer99}.
They showed that when each link was resonant,  the dark state of STIRAP \cite{STIRAP} could be modified to a higher-order trapping state,
 becoming an approximate constant of motion even for small pulse areas.
This state adiabatically rotates, in Hilbert space, from the initial to the target state.
This adiabatic motion  leads to efficient population transfer, though at the expense of placing some population into the decaying atomic state.

Recently a three-state loop was shown to occur in physical processes where the free-space symmetry is broken, as it is in chiral systems \cite{Shapiro,Liu}.
Such quantum systems occur in  left- and right-handed chiral molecules \cite{Shapiro}, or in ``artificial atoms''.
Loop linkages  amongst discrete quantum states can also occur in superconducting quantum circuits \cite{Liu},
 and in modeling entangled atoms coupled through an optical cavity \cite{Metz06}.

We here consider loops that have less stringent constraints on the frequencies, although some do exist.
We shall show in the following that it is possible, under appropriate conditions (including three-photon resonance), to break the loop into a chain.
A further transformation of the basis states can then convert the linkage pattern into a pair of coupled states and a spectator state.
Alternatively, if the population resides initially in the middle state of the chain, the system has the dynamics of the vee linkage.

The required initial transformation, converting the loop into a simple chain, is taken to be a Householder reflection (HR) of the Hamiltonian matrix \cite{Householder}.
Such matrix manipulations, commonplace in works dealing with linear algebra \cite{Householder2}, have recently been applied to quantum-state manipulations \cite{Kyo06,Iva06,Kyo07,Iva07}.

When acting upon an \textit{arbitrary} square matrix a suitable sequence of HRs produces an upper-diagonal (or lower-diagonal) matrix.
When acting upon a \textit{unitary} matrix, such a sequence produces a diagonal matrix, with phase factors on the diagonal.
This property has been used for decomposition, and therefore synthesis, of arbitrary preselected propagators in multistate systems \cite{Kyo06,Iva06,Kyo07,Iva07}.
We show here that, when utilised for a change of basis in Hilbert space, a suitable HR (or a sequence of HRs) can cast a (hermitian) Hamiltonian into a tridiagonal form.
This tridiagonalization implies the replacement of a general linkage pattern (for example, each state interacting with any other state)
 with an effective chainwise-connected system where only nearest-neighbour interactions are present.
We here apply this tridiagonalization to the simplest nontrivial multistate system -- a three-state loop system -- and demonstrate its potential applications,
 with examples ranging from effective chain breaking and novel spectator states to hidden two-photon resonances.

\section{The loop RWA Hamiltonian}

We consider three fields, labelled pump ($P$), Stokes ($S$), and control ($C$),
\be
\mathbf{E}_k(t) = \hat{\mathbf{e}}_k \, \mathcal{E}_k(t) \, \cos(\omega_k t + \phi_k), \qquad (k = P, S, C).
\ee
The three carrier frequencies $\omega_k$ can be chosen arbitrarily, so long as they fulfill the three-photon resonance condition (Fig. \ref{fig-loop3})
\be
\omega_C - \omega_P + \omega_S = 0.
\label{eqn-3phot}
\ee
This constraint is necessary for the application of the rotating-wave approximation (RWA) \cite{All75,Shore}.
However, at the outset we impose no constraints on the single-photon detunings,
\be
\hbar \Delta_P
\equiv E_2-E_1
-\hbar \omega_P,\qquad \hbar \Delta_S\equiv E_2-E_3-\hbar \omega_S.
\ee

We introduce probability amplitudes $C_n(t)$ in the usual rotating Hilbert-space coordinates $\state{n}(t)$,
\be
\Psi(t) = \exp(- i \zeta_0 t ) \left[ C_1(t) \state{1} + C_2(t) \state{2}(t) + C_3(t) \state{3} (t) \right],
\ee
where the rotations originate with field carrier frequencies,
$\state{2}(t) \equiv \exp(-i \omega_P t) \state{2}$ and $\state{3}(t) \equiv \exp(-i\omega_C t ) \state{3}$.
From the time-dependent Schr\"{o}dinger equation we obtain three coupled equations
\be
  \frac{d}{dt} \C(t) =  -i \W(t) \C(t),
\ee
where $\C(t)\equiv [C_1(t),C_2(t),C_3(t)]^{T}$ is a three-component column vector of probability amplitudes and $\hbar \W(t)$ is the slowly varying RWA Hamiltonian.
We take the overall phase factor $\zeta_{0}$ to nullify the first diagonal element of the RWA Hamiltonian matrix; it then has the structure
\be \label{H-loop}
\mathsf{W}(t)=\frac{_1}{^{2}}\left[
\begin{array}{ccc}
0 & \Omega_P(t)e^{i\phi_P} & \Omega_C(t)
 \\
\Omega_P(t)e^{-i\phi_P} & 2\Delta_2 & \Omega_S(t)e^{i\phi_S}
\\
  \Omega_C(t) & \Omega_S(t)e^{-i\phi_S} & 2\Delta_3%
\end{array}
\right] . \ee where the interactions are parameterized by slowly
varying real-valued Rabi frequencies $\Omega_k(t)$, for $k =
P,S,C$. For simplicity and without loss of generality the $C$
field is assumed real ($\phi_C=0$); then $\phi_P$ and $\phi_S$
represent the phase differences between the $P$ and $S$ fields,
respectively, and the $C$ field. The cumulative detunings are \be
  \Delta_2=\Delta_P,\quad \Delta_3 = \Delta_P-\Delta_S.
\ee

\section{The Householder reflection}

We seek a unitary transformation of the Hilbert-space basis states that will first replace the loop with a three-state chain.
As we will show, the desired result can be produced by a Householder reflection acting upon the RWA Hamiltonian \cite{Iva06}.

An $N$-dimensional Householder reflection is defined as the operator
\be
 \R = \mathsf{I} - 2\left\vert v\right\rangle \left\langle v\right\vert ,
\ee
where $\mathsf{I}$ is the identity operator and $\left\vert v\right\rangle $ is an $N$-dimensional normalized complex column vector.
The Householder operator $ \R $ is Hermitian and unitary, $ \R = \R ^{-1}= \R ^{\dagger }$, hence $ \R $ is involutary, $ \R ^{2}=\mathsf{I}$.
The  transformation is a reflection, so $\det  \R =-1$.
If the vector $\left\vert v\right\rangle $ is real, the Householder reflection has a simple geometric interpretation:
 it is a reflection with respect to an $(N-1)$-dimensional plane normal to the vector $\left\vert v\right\rangle $.

The Householder reflection, acting upon an arbitrary $N $-dimensional matrix, uses $N-1$  operations to produce  an upper or lower triangular matrix.
This behavior makes the Householder reflection a powerful tool for many applications in classical data analysis, e.g., in
solving systems of linear algebraic equations, finding eigenvalues of high-dimensional matrices, least-squares optimization, QR decomposition, etc. \cite{Householder2}.
For us, the reflection serves to transform the Hamiltonian from a full matrix to one that lacks one interaction -- it breaks the loop into a chain.

\begin{figure}[tb]
\epsfig{width=75mm,file=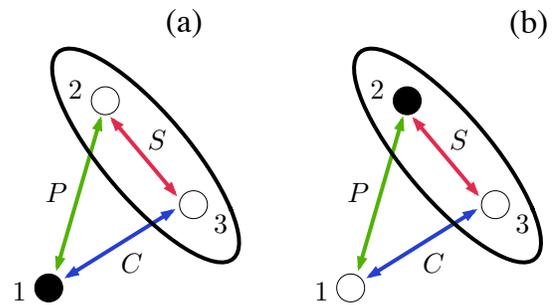} \caption{(Color online) The
Householder reflection leaves state $\state{1}$ unchanged. Initial
population might be (a) in this state, or (b) in one of the
altered states. Relative energies of the original bare states are
not relevant, only the couplings. } \label{fig-Househ}
\end{figure}

The three-state system offers three basis vectors with which to define a Householder reflection.
Because of the symmetry of the loop system it is only necessary to consider one of these; the effect of others can be examined by a permutation of state labels.
We shall take state $\state{1}$ as a fixed coordinate, within the plane of the reflection, and introduce  an alteration of the Hilbert subspace spanned by the remaining unit vectors $\state{2}(t)$ and  $\state{3}(t)$.
Figure \ref{fig-Househ} illustrates the connection of the reflection with the basis states, and the possible choices of initial state.

With this choice the Householder vector reads
\be\label{Householder vector}
|v\rangle = \left[ 0,\sin (\theta/2) e^{-i\phi_{P}}, \cos (\theta/2) \right] ^{T},
\ee
and the matrix representation of the Householder reflection is
\be
\R \t  =\left[\begin{array}{ccc}
1 & 0 & 0 \\
0 & \cos \theta  \t   & e^{-i\phi_P}\sin \theta  \t   \\
0 & e^{i\phi_P}\sin \theta  \t   & -\cos \theta  \t  %
\end{array}\right].
\ee
The angle $\theta$, defined by the equation
\be \label{theta}
\tan \theta  \t \equiv \frac{\Omega_C \t  }{\Omega_P \t  },
\ee
is twice the angle from the mirror normal to state $\state{2}$ (i.e. the twist of the mirror about the $\state{1}$ axis).
Hereafter we omit explicit mention of time dependences; all Rabi frequencies are to be considered slowly varying in time, as is the Householder reflection $\R$ and the angle $\theta$.

The connection between the probability amplitudes $ \tC  \t  $ in the Householder basis and the amplitudes $\C \t  $ in the original (bare) basis is
\be
 \tC \t  = \R \t \C \t  .
\ee
The transformed equation of motion reads
\be
 \frac{d}{dt} \tC  \t  = -i \tW  \t   \tC  \t,
\ee
where the Householder Hamiltonian is $\tW  \t  = \R  \t   \W \t   \R  \t  -i \R  \t   \dot{ \R } \t$, with an overdot denoting a time derivative. Explicitly,
\be\label{HH}
 \tW  \t  =\tfrac{1}{2}\left[\begin{array}{ccc}
0 & \tV_P  \t   & \ 0
\\
\tV_P^*  \t   & 2\tDelta_2  \t   & \tV_S  \t -2ie^{-i\phi_{p}}\dot{\theta }  \t
\\
\ 0 & \quad \tV_S^*  \t  + 2ie^{i\phi_{p}}\dot{\theta } \t   \quad & 2\tDelta_3  \t
\end{array}\right] ,
\ee
with effective detunings
\bse\label{HDelta}\bea
\tDelta_2 \t &=& \frac{\Delta_3 \Omega_C \t ^2 + \Delta_2 \Omega_P \t ^2 +\Omega_P \t \Omega_C \t \Omega_S \t \cos \left( \phi_P+\phi_S\right) }{\Omega \t ^2},\label{HDelta2}
 \\
\tDelta_3 \t &=& \frac{\Delta_2 \Omega_C \t ^2 + \Delta_3 \Omega_P \t ^2 -\Omega_P \t \Omega_C \t \Omega_S \t \cos \left( \phi_P+\phi_S\right) }{\Omega \t ^2},\label{HDelta3}
\eea\ese
and effective couplings
\bse\label{HOmega}\bea
\tV_P  \t   &=&e^{i\phi_{p}}\Omega \t ,\label{HOmegaP}
 \\
\tV_S  \t   &=&\frac 1{\Omega \t  ^2} \Big[ 2e^{-i\phi_P} \left( \Delta_2 -\Delta_3  \right) \Omega_P \t  \Omega_C \t  \notag\\
&& + \left.\left(e^{-2i(\phi_P+\phi_S) }\Omega_C \t  ^2 -\Omega_P \t  ^2 \right) e^{i\phi_S} \Omega_S \t \right],\label{HOmegaS}
\eea\ese
with $\Omega \t  \equiv  \sqrt{\Omega_P \t  ^2+ \Omega_C \t  ^2}$.
All of these elements  acquire time dependence from the pulses, though that is not shown explicitly.

The Hamiltonian in the Householder basis is that of a simple chain, $\state{1}\leftrightarrow \tstate{2}\leftrightarrow \tstate{3}$.
By design the Householder reflection places  the original two interactions of state $\state{1}$ into a single effective interaction with a new superposition state $\tstate{2}$.
This state, in turn, has an interaction with the other terminal state of the chain $\tstate{3}$, also a superposition state.
The new \emph{Householder states} $\tpsi_{n} \t$ are superpositions of the original basis states $\state{n} \t$,
\bse\label{Householder states}
\bea
\tstate{1}  &=&\state{1}, \label{Householder state 1}\\
\tstate{2}  \t   &=&\cos \theta  \t  \, \state{2} \t + e^{-i\phi_P}\sin \theta  \t  \, \state{3} \t  , \label{Householder state 2}\\
\tstate{3}  \t   &=&e^{i\phi_P}\sin \theta  \t  \, \state{2} \t - \cos\theta  \t   \, \state{3} \t  \label{Householder state 3}.
\eea \ese
When the initial population resides entirely in state $\state{1}$, this chain is equivalent to a lambda or ladder system.
When the initial population occurs in state $\tstate{2}$ it is a generalization of the vee linkage.
The chain Hamiltonian \eqref{HH} in the Householder representation is conceptually simpler than the original loop Hamiltonian \eqref{H-loop}
 for it allows only for nearest-neighbour interactions; the resulting chain linkage is easier to understand and treat analytically by a variety of exact or approximate approaches.
The inherent interference in the loop system is now imprinted onto the Householder transformation and  is absent in the Householder chain.
Moreover, this tranformation allows one to use the considerable literature available on chainwise-connected three-state systems.

\section{Special cases} 

In the remainder of this paper we consider special cases of the Householder Hamiltonian,
 obtained when we constrain the various pulse parameters, which lead to simplification of the resulting Hamiltonian matrix.
Two simplifications are particularly interesting:
 (i) breaking the three-state Householder chain $\state{1}\leftrightarrow \tstate{2}\leftrightarrow \tstate{3}$  into a two-state system and a spectator state,
 and (ii) two-photon resonance in the Householder basis.
We shall identify conditions, and deduce implications, for these important special cases.

\subsection{Effective two-state system and spectator state}

Under appropriate conditions the three-state Householder chain $\state{1}\leftrightarrow \tstate{2}\leftrightarrow \tstate{3}$ breaks into two coupled states and a spectator state.
This occurs whenever one of the Householder linkages vanishes.
The vanishing of $\tV_P$ requires that both $\Omega_P$ and $\Omega_C$ vanish, which is trivial and uninteresting.
We hence assume the null linkage  to be the coupling between states $\tstate{2}$  and $\tstate{3}$,
\be\label{H23=0}
 \tV_S \t + 2i e^{-i\phi_p}\dot{\theta} \t = 0.
\ee
Under this condition state $\tstate{3}$, Eq.~\eqref{Householder state 3}, decouples from the other two states and becomes a \textit{spectator state}:
 its population is trapped, within a subspace of the full Hilbert space.
The population distribution between states $\state{2}$ and $\state{3}$ may change, but in a manner that conserves the population of the spectator state \eqref{Householder state 3}.

\subsubsection{Conditions for chain breaking}

One possible solution to Eq.~\eqref{H23=0} reads
\bse\label{breaking conditions}\bea
\Delta_2 &=&\Delta_3, \label{beaking condition 1}\\
\phi_P &=& -\phi_S-\frac{\pi }{2}, \label{breaking condition 2}\\
\Omega_S \t &=&-2\dot{\theta } \t. \label{breaking condition 3}
\eea\ese
The latter condition imposes a strict constraint on the pulse shape.
Given $\Omega_P$ and $\Omega_C$, which determine $\dot\theta$ through Eq.~\eqref{theta}, condition \eqref{breaking condition 3} determines both the shape and the magnitude of $\Omega_S$.

Another possible solution to Eq.~\eqref{H23=0} emerges when the $P$ and $C$ pulses have the same time dependence, say $f(t)$,
 $\Omega_P (t) =\Omega_{P0} f(t)$, $\Omega_C (t) = \Omega_{C0} f(t)$,
 while the $S$ pulse could differ, $\Omega_S (t)  =\Omega_{S0} g(t)$.
Then $\dot{\theta} = 0$ and  condition \eqref{H23=0} becomes $\tV_S \t =0$.
This condition can be satisfied in several ways, cf. Eq.~\eqref{HOmegaS}.
A simple realization for that condition occurs with the choice
\bse\label{breaking conditions b}\bea
\Delta_2 &=&\Delta_3, \label{breaking condition 1b}\\
\phi_P&=&-\phi_S, \label{breaking condition 2b}\\
\Omega_C \t &=&\Omega_P \t . \label{breaking condition 3b}
\eea\ese
Then $\theta=\pi/4$ and the spectator state reads
\be\label{dark state in case of no adiabatic coupling}
\tstate{3}  \t = \tfrac{1}{\sqrt{2}} \left(  e^{-i\phi_S}\state{2} - \state{3} \right).
\ee

\subsubsection{Analytical three-state solutions}

The dynamics of the two coupled Householder states $\state{1}$ and $\tstate{2}$, coupled by the interaction $\tV_P \t  $,
 offer other interesting possibilities.
For the two-state system $\state{1}\leftrightarrow\tstate{2}$, analytic solutions may be possible; these are known for many examples of pulse and detuning time dependences.
Hence, by writing down the propagator in the Householder basis for a known two-state analytical solution,
 and by using the transformation back to the original basis by the Householder reflection $\R(t)$, one can write down a number of analytic solutions for the three-state loop system.
These would generalize the similar analytical solutions for a $\Lambda$ system \cite{Vitanov98}.

\subsubsection{Population initially in state $\state{1}$}

If only state $\state{1}$ is initially populated then  the dynamics remains confined within the effective two-state system $\state{1}\leftrightarrow\tstate{2}$.
In this two-state system we can enforce complete population return to state $\state{1}$,
 complete population inversion to state $\tstate{2}$ or create a superposition of states $\state{1}$ and $\tstate{2}$.

\textit{Complete population transfer} from state $\state{1}$ to the Householder state $\tstate{2}$ can be produced by a resonant $\pi$-pulse \cite{Shore},
 by adiabatic level-crossing adiabatic passage \cite{ARPC},
 or by a variety of novel more sophisticated techniques \cite{SCRAP,RIBAP,SAP,deltafn}.
Viewed in the original basis, the system ends up in a superposition of $\state{2}$ and $\state{3}$,
\be
\cos \theta \t \, \state{2} \t +e^{-i\phi_P}\sin \theta \t \state{3} \t,
\ee
with the angle $\theta$ given by (\ref{theta});
 thus the superposition is fully controlled by the ratio of $\Omega_C \t $ and $\Omega_P \t $ and has a relative phase $\phi_P $.

\begin{figure}[tb]
\epsfig{width=75mm,file=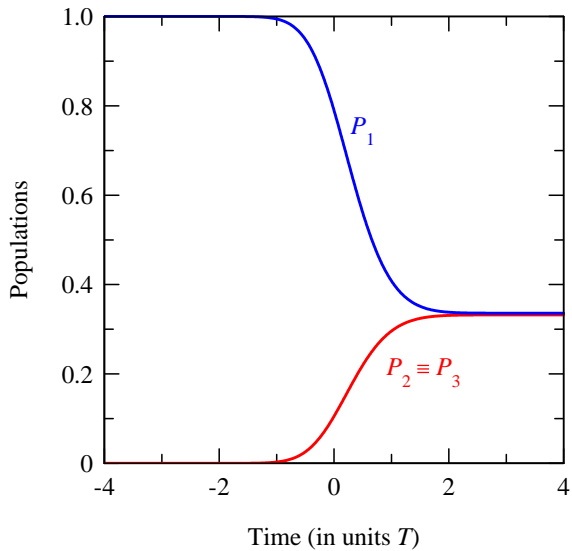} \caption{(Color online) Creation
of an equal superposition of states $\state{1}$, $\state{2}$ and
$\state{3}$
 for Gaussian pulses: $\Omega_P(t) = \Omega_{P0}\ e^{-t^2/T^2}$, $\Omega_C(t) = \Omega_{C0}\ e^{-t^2/T^2}$,
 $\Delta_2(t)=\Delta_3(t)=-\Omega_S(t)/2$, $\Omega_S(t)=\Omega_{S0}\ e^{-(t-\tau)^2}$,
 with the following parameters $\Omega_{P0}=\Omega_{C0}=0.76/T$, $\Omega _{S0}=1/T$, $\tau=0.5T$.}
\label{superposition1}
\end{figure}

A predetermined superposition of states $\state{1}$ and $\tstate{2}$ can be created by resonant fractional-$\pi$ pulses,
 or by modifications of adiabatic-passage techniques, for example, half-SCRAP \cite{Half-SCRAP} and two-state STIRAP \cite{2s-STIRAP}.
Such techniques allow, for instance, the creation of an arbitrary predetermined maximally coherent superposition of the three states $\state{1}$, $\state{2}$, and $\state{3}$.
For example, one can create a maximally coherent superposition
 using  fractional-$\pi $ pulses that obey  conditions (\ref{breaking conditions b})
 and which are resonant in the Householder basis ($\Delta_2(t) =\Delta_3(t) =-\Omega_S(t)/2 $).
Such an example is demonstrated in Fig.~\ref{superposition1}.

\subsubsection{Population initially in state $\state{2}$}

\begin{figure}[tb]
\epsfig{width=75mm,file=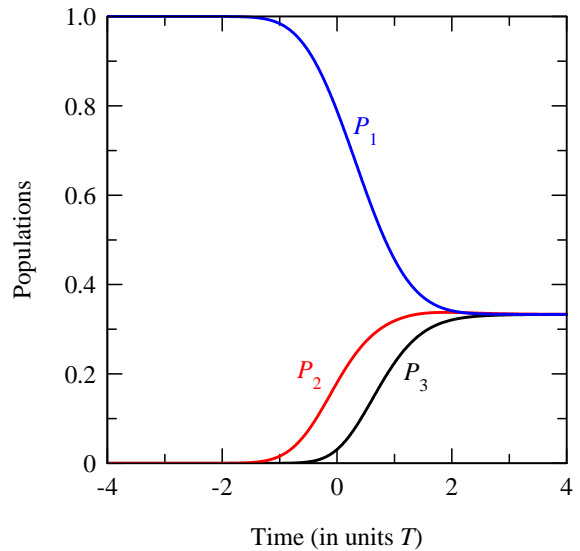} \caption{(Color online) Creation
of an equal superposition of states $\state{1}$, $\state{2}$ and
$\state{3}$
 for Gaussian pulses: $\Omega_P(t) = \Omega_{0}\ e^{-(t+\tau)^2/T^2} + \Omega_{0} \cos\alpha\ e^{-(t-\tau)^2/T^2}$,
 $\Omega_C(t) = \Omega_{0} \sin\alpha\ e^{-(t-\tau)^2/T^2}$,
 $\Delta_2(t)=\Delta_3(t)=0$, $\Omega_S(t)=-2\dot\theta$,
 with the following parameters $\alpha=\pi/4$, $\Omega_0=0.567/T$, $\tau=0.5T$.}
\label{superposition2}
\end{figure}

Let us assume now that it is state $\state{2}$ that is populated initially.
 (The symmetric case of state $\state{3}$ initially populated is just a matter of relabelling the states.)
If the $C$ pulse precedes the $P$ pulse, then we are in the dark state (\ref{dark state}) and this is a situation similar to STIRAP, then there will occur complete population transfer to state $\state{3}$.
The resonant case of this process was discussed and explained earlier \cite{Unanyan97,Fleischhauer99}.
If we are in state $\state{2}$ and we apply the pulse in the intuitive order (the $P$ pulse precedes the $C$ pulse), then we are in the bright state and
 depending on the pulses we can have complete population transfer to state $\state{1}$, or end up in a superposition of states $\state{1}$, $\state{2}$ and $\state{3}$.
For example, as illustrated in Fig. \ref{superposition2}, one can use fractional-$\pi$ pulses to create an equal superposition of states $\state{1}$, $\state{2}$ and $\state{3}$.

\subsection{Effective two-photon resonance}

We assume now that the $P$ and $C$ pulses have the same time dependence and consider a resonance condition between states $\state{1}$ and $\tstate{3}$,
\be\label{resonance condition}
\tDelta_3  \t  =0.
\ee
The resulting Householder Hamiltonian is exactly that of the lambda linkage on two-photon resonance used for STIRAP \cite{STIRAP}.
The traditional dark state of the STIRAP process appears here as
\bea
\Phi_{D} \t   &=&\cos \varphi  \t  \,\state{1} - \sin \varphi \t  \,\tstate{3}  \t
\notag \\
&=&\cos \varphi  \t  \,\state{1} -e^{i\phi_P}\sin \varphi  \t  \,\sin\theta \,\state{2} \t + \sin \varphi  \t \,\cos \theta \state{3} \t  ,
 \label{dark state}
\eea
where $\tan \varphi \t = \tV_P \t / \tV_S \t$.
The state $\Phi_{D} \t  $ is a spectator (or population-trapping) state because it is not affected by the specified radiation, but it has components of all three of the original basis states.
One can use this new kind of spectator state, with the traditional STIRAP pulse sequence  of $\tV_S  \t  $ preceding $\tV_P  \t$,
 to move the initial population from state $\state{1}$ to a superposition of state $\tstate{2} \t$ and state $\tstate{3} \t$.
The superposition is controlled by the ratio of $\Omega_C \t $ and $\Omega_P \t $ and has the phase $\phi_P$.

Condition \eqref{resonance condition} can always be satisfied for an appropriate choice of the (time-dependent) detuning $\Delta_2(t)$ [or $\Delta_3(t)$].
However, the specific time dependence, although possible in principle, might be complicated and difficult  to produce experimentally.
Condition \eqref{resonance condition} can be satisfied with constant detunings when the $P$ and $C$ pulses share the same time dependence: $\Omega_P (t) =\Omega_P f(t)$ and $\Omega_C (t) = \Omega_C f(t)$.
Then the mixing angle $\theta$ is constant [see Eq.~\eqref{theta}] and $\dot\theta=0$.
Two options provide the needed pulses:

(i) The conditions
\bse\label{2PR conditions}\bea
&&\phi_P+\phi_S = \pi /2, \label{phase condition}\\
&& \Delta_3 = -\Delta_2\frac{\Omega_C^2}{\Omega_P^2}\label{detuning condition}
\eea\ese
hold. Then the $S$ field can be arbitrary and both detunings $\Delta_2$ and $\Delta_3$ can be constant.

(ii) The $S$ field is constant. Then condition \eqref{resonance condition} can be fulfilled for constant detunings that obey the relation
\be\label{detuning condition 2}
\Delta_3 = -\Delta_2\frac{\Omega_C^2}{\Omega_P^2} + \frac{\Omega_C}{\Omega_P}\Omega_S \cos(\phi_P+\phi_S).
\ee

Therefore, the usual two-photon resonance condition, necessary for the emergence of a spectator (dark) state in the original basis,
 is replaced by a condition for the two-photon detuning $\Delta_3$:
 either (i)  Eq.~\eqref{detuning condition}, for arbitrary $S$ field but with the phase relation \eqref{phase condition},
 or (ii) Eq.~\eqref{detuning condition 2}, for constant $S$ field.

\begin{figure}[tb]
\epsfig{width=75mm,file=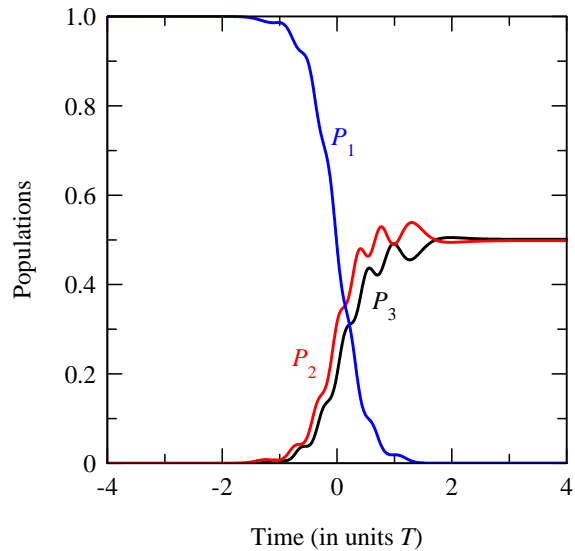} \caption{(Color online) Creation
of an equal superposition of states $\state{2}$ and $\state{3}$,
with the following couplings and detunings:
 $\Omega_P(t) = \Omega_0 \cos\theta\ e^{-(t-\tau)^2/T^2}$,
 $\Omega_C(t) = \Omega_0 \sin\theta\ e^{-(t-\tau)^2/T^2}$,
 $\Omega_S(t) = \Omega_0\ e^{-(t+\tau)^2/T^2} $, $\Delta_2(t) = \Delta_3(t) = 0$,
 where the parameters are $\theta = \pi /4$, $\Omega_0=30/T$, $\tau=0.5T$.}
\label{superposition3}
\end{figure}

If we now start  initially in state $\state{1}$ and apply the $S$ pulse before the $P$ pulse, then the following superposition
is formed
\be -ie^{-i\phi_S}  \,\sin \theta \,\state{2} \t + \cos \theta \state{3} \t. \ee
The superposition characteristics are fixed by the $S$-field phase and the angle $\theta$ defined by Eq.~\eqref{theta}.
Figure \ref{superposition3} illustrates how,  starting in state $\state{1}$ and applying the $S$ pulse before the $P$ pulse we obtain an equal superposition of  $\state{2}$ and $\state{3}$.

\section{Reduction of arbitrary $N$-dimensional quantum systems to chains}

The Householder transformation introduced here for a three-state loop system is readily extended to a general $N$-state quantum system with arbitrary linkages,
 even in the most general case when each state connects to any other state.
A suitable sequence of at most $N-2$ Householder transformations can cast the Hamiltonian, which is a hermitian matrix, into a \textit{tridiagonal} form.
In this sequence the Householder vector for the $n$th reflection is chosen as
\be
|v_{n}\rangle =\frac{|x_{n}\rangle -\left\vert x_{n}\right\vert
|e_{n+1}\rangle }{\left\vert |x_{n}\rangle -\left\vert x_{n}\right\vert |e_{n+1}\rangle \right\vert },
\ee
where $|e_{n+1}\rangle $ is a unit vector that defines the $(n+1)$-st axis, i.e. its components are zero except a unity at the $(n+1)$st place,
Here $|x_{n}\rangle $ is the $n$th column vector of the transformed Hamiltonian after the $n$th step.

The tridiagonalization of the Hamiltonian implies that in the Householder basis, each of the new basis states is connected only to its nearest neighbor states, thus forming a \textit{chainwise} linkage pattern.
The chain is conceptually simpler, and analytically easier, to treat, with a variety of exact and approximate approaches available in the literature.

\Black
\section{Conclusions and outlook}

A two-parameter Householder reflection can break the loop-linkage pattern of a three-state system, providing instead a simple chain.
For the three-state system the result can appear either as a lambda system (with initial population at one end of the chain) or as a vee system (with initial population in the middle state of the chain).
In either case the system can be transformed further into a pair of coupled states and a spectator state, within which population remains trapped.
This is a new kind of spectator state involving all three basis states; it contrasts with the conventional dark states that have no excited-state component.

These results hold  intrinsic interest because the three-state loop is the simplest discrete-state quantum system in which nontrivial interference occurs.
The present solutions may therefore offer opportunities for manipulating the quantum states of such systems.

Our objective in this paper has been to introduce this important novel transformation, and with it to show  that a loop system is equivalent to a chain system.
The presented examples of the uses of the Householder transformation in a three-state loop system, being by no means exhaustive,
 have demonstrated a number of potential applications based on analytical approaches, ranging from hidden chain breaking and spectator states to hidden two-photon resonances and analytic solutions.
These allow one to establish generic features of the interaction dynamics and engineer interactions that can produce various superposition states at will.

The results in this work for three-state systems are readily extended to $N$-state systems.
More general sequences of Householder reflections can replace there arbitrary complicated linkages with simple chain linkages.
Hence an $N$-state system wherein each state is coupled to any other state  can be reduced to an equivalent chain system with nearest-neighbor interactions only.
Then one can apply various available analytical approaches to the Householder chain to reveal interesting novel features of the multistate dynamics,
 including Hilbert space factorization, hidden spectator states and ensuing dynamical invariants.


\acknowledgments

This work has been supported by the EU ToK project CAMEL (Grant No. MTKD-CT-2004-014427), the EU RTN project EMALI (Grant No. MRTN-CT-2006-035369),
 and the Bulgarian National Science Fund Grants No. WU-205/06 and No. WU-2517/07.


\def\bibttl#1{}

\end{document}